\documentclass[prl, reprint, amsmath, amssymb, longbibliography]{revtex4-1}

\usepackage{hyperref}
\usepackage{graphicx}
\usepackage{dcolumn}
\usepackage{textcomp}
\usepackage{color}
\usepackage[english]{babel}
\usepackage{amsmath}
\usepackage{amssymb}
\usepackage{bm}
\usepackage{upgreek}
\usepackage{lineno}
\usepackage{xcolor}
\usepackage[defaultcolor=red,final]{changes}
\pdfoutput=1

\usepackage{epstopdf}

\begin{document}

\title{Optimal transfer functions for bandwidth-limited imaging}

\author{Sjoerd Stallinga}
\email{s.stallinga@tudelft.nl}
\affiliation{Department of Imaging Physics, Delft University of Technology, 2628 CJ Delft, The Netherlands}

\author{\added{Niels Radmacher}}
\email{niels.radmacher@uni-goettingen.de}
\affiliation{\added{III. Institute of Physics, Georg-August University, 37077 G\"ottingen, Germany}}

\author{Antoine Delon}
\email{antoine.delon@univ-grenoble-alpes.fr}
\affiliation{Universit\'e Grenoble Alpes, CNRS, LIPhy, 38000 Grenoble, France}

\author{J\"org Enderlein}
\email{jenderl@gwdg.de}
\affiliation{III. Institute of Physics, Georg-August University, 37077 G\"ottingen, Germany}
\affiliation{Cluster of Excellence ``Multiscale Bioimaging: from Molecular Machines to Networks of Excitable Cells'' (MBExC), Georg August University, 37077 G\"ottingen, Germany}

\date{\today}


\begin{abstract}
One of the fundamental limits of classical optical microscopy is the diffraction limit of optical resolution. It results from the finite bandwidth of the optical transfer function (or OTF) of an optical microscope, which restricts the maximum spatial frequencies that are transmitted by a microscope. However, given the frequency support of the OTF, which is fully determined by the used optical hardware, an open and unsolved question is what is the optimal amplitude and phase distribution of spatial frequencies across this support that delivers the ``sharpest'' possible image. In this paper, we will answer this question and present a general rule how to find the optimal OTF for any given imaging system. We discuss our result in the context of optical microscopy, by considering in particular the cases of wide-field microscopy, confocal Image Scanning Microscopy (ISM), 4pi microscopy, and Structured Illumination Microscopy (SIM). Our results are important for finding optimal deconvolution algorithms for microscopy images, and we demonstrate this experimentally on the example of ISM. They can also serve as a guideline for designing optical systems that deliver best possible images, and can be easily generalized to non-optical imaging such as telescopic imaging, ultrasound imaging, or magnetic resonance imaging.
\end{abstract}

\maketitle

\section{Introduction}

Conventional optical microscopes have a resolution limit due to the wave nature and resulting diffraction of light. This is embodied by the fact that a microscope transmits only spatial frequencies of a sample up to some maximum values. The total set of all transmitted frequencies comprises the support of the so-called Optical Transfer Function (OTF). The OTF is the Fourier transform of the Point Spread Function (PSF), the image of an ideal infinitesimally small point emitter. 
Recorded data are often deconvolved using a model of image formation for the given imaging system \cite{Richardson1972RLdecon,Lucy1974RLdecon,vankempen1997quantitative,verveer1999comparison,wallace2001deconvolution,sibarita2005deconvolution}. The goals of such a deconvolution can be several: (i) correcting optical aberrations of the imaging system, (ii) maximizing spatial resolution and contrast, or (iii) minimizing out-of-focal plane contributions. An appropriate model of the imaging system is also required for non-direct imaging methods, that need an image reconstruction procedure to arrive at the final image. For example, image reconstruction in Structured Illumination Microscopy (SIM) \cite{heintzmann1999sim,gustafsson2000sim,Gustafsson_Shao_Carlton_Wang_Golubovskaya_Cande_Agard_Sedat_2008,Kraus_Miron_Demmerle_Chitiashvili_Budco_Alle_Matsuda_Leonhardt_Schermelleh_Markaki_2017,heintzmann2017simreview} requires finding an optimal apodization function. In any case, an optimal deconvolution or image reconstruction needs a valid model of the ``ideal'' OTF (or, equivalently PSF) for a given imaging system, which would yield the best possible image of a sample.

The goals of this paper are threefold. First, we want to answer the fundamental question in optical imaging what the ideal OTF is given the support in spatial frequency space, and how such an ideal OTF can be found from first principles. Second, we want to illustrate this concept to find the ideal OTF for a range of microscopic imaging modalities. Third, we wish to demonstrate a practical application of the concept of an ideal OTF in linear deconvolution.  

\section{Theoretical Foundation} 

\added{In what follows, we will focus on image formation in fluorescence microscopy}. Let us start by considering the image formation of a general optical microscope. To be as general as possible, we consider three-dimensional image formation, either by taking consecutive images at different focal planes along the optical axis with a wide-field microscope (with or without structured illumination), or by three-dimensional point scanning with a laser scanning microscope (e.g. conventional confocal microscope, image scanning microscope, STED microscope). For an optical system that is isoplanatic along all directions \cite{welford1976vi}, the final 3D image $I(\mathbf{r})$ is related to the sample function $S(\mathbf{r})$ (distribution density of fluorescent emitters) by a linear convolution of the form

\begin{equation} 
\begin{split}
I(\mathbf{r}) = \int d\mathbf{r}' U(\mathbf{r}-\mathbf{r}') S(\mathbf{r}') \Leftrightarrow \tilde{I}(\mathbf{k}) = \tilde{U}(\mathbf{k}) \tilde{S}(\mathbf{k}), 
\end{split} 
\end{equation}

\noindent where $\mathbf{r}$ and $\mathbf{k}$ are the coordinates in real and Fourier space, respectively, and a tilde above a symbol denotes its Fourier transform. Here, $U(\mathbf{r})$ is the PSF, and $\tilde{U}(\mathbf{k})$ is its Fourier transform, the OTF. Along all directions $\hat{\mathbf{n}}$, there exists a maximum value $k_{max}(\hat{\mathbf{n}})$ beyond which the OTF is zero, which means that no spatial frequencies of the sample distribution $\tilde{S}(\mathbf{k})$ beyond values $k_{max}(\hat{\mathbf{n}}) \hat{\mathbf{n}}$ are transmitted by the microscope (final support of the OTF). This fundamentally limits the achievable spatial resolution along direction $\hat{\mathbf{n}}$ to $2\pi/k_{max}$. We will denote by $\mathcal{M}$ the set of all points of the support of the OTF in Fourier space, i.e.

\begin{equation} \begin{split}
\mathcal{M} = \{ \mathbf{k}\, \big\vert\, \vert\tilde{U}(\mathbf{k})\vert >0 \}.
\end{split} \end{equation}

\noindent This support $\mathcal{M}$ is solely determined by the microscope's hardware properties, such as the numerical aperture of its objective, the size of the confocal aperture and magnification at the aperture's position (for a confocal microscope), the spatial modulation period of the excitation intensity (for a structured illumination microscope), or the intensity of the stimulated emission laser (for a STED microscope). But given this support $\mathcal{M}$, the question is what is the distribution of complex-valued amplitudes over $\mathcal{M}$ that yields the best possible PSF. 

At this point it makes sense to make more explicit what we understand to be the ideal OTF and PSF. Our position is that an ideal PSF of an incoherently imaging fluorescence microscope should be (i) positive everywhere ($U(\mathbf{r})\geq 0$), as this avoids edge ringing and negative pixel artifacts in linear deconvolution and image filtering, as well as being consistent with the physical requirement of having non-negative image signals, (ii) as sharp as possible, which can be quantified by a maximum OTF summed over the OTF support, or equivalently a maximum PSF peak value

\begin{equation} \begin{split}
U(0) = \int_{\mathcal{M}}\frac{d\mathbf{k}}{(2\pi)^D}\tilde{U}(\mathbf{k}),
\label{eq:sharpnessmetric}
\end{split} \end{equation}

\noindent where $D$ is the dimensionality of $\mathcal{M}$. \added{The advantage of a sharpness metric that is based on the OTF is that is independent of any specific object features, in contrast to e.g. the steepness of imaged edges.}.

\added{There is a large interdependency between the above two requirements. For example, a gain in contrast and overall sharpness may be achieved by having a relatively high OTF for large spatial frequencies, as this will emphasize the representation of small scale features in the image. It appears, however, that an unbalanced OTF, with comparable or even higher values for larger spatial frequencies than for lower spatial frequencies, gives rise to violations of the positivity constraint for the PSF. A metric to quantify sharpness should therefore take into account the OTF at all spatial frequencies, a straightforward way to do this is to simply take the average over all spatial frequencies in the OTF support.} The quest for a best possible OTF giving rise to a positive PSF has already been addressed by Lukosz, who has derived upper bounds for such an OTF \cite{lukosz1962bounda,lukosz1962boundb}. The concept of the Lukosz bound has also been applied in image reconstruction for SIM \cite{righolt2013image,righolt2014three}. Here, we improve on these results by proposing an exact recipe for obtaining the ideal OTF given the support $\mathcal{M}$.

Let us start with the simplest case of an infinitely extended support in one dimension. This is a trivial case with the well-known answer that a constant OTF, $\tilde{U}(\mathbf{k})=\mathrm{const}$, yields the most narrow PSF, namely a delta function. In that case, the image $I(\mathbf{r})$ and the sample function $S(\mathbf{r})$ become identical. For the one-dimensional case with a limited support $\left\vert k\right\vert<k_{max}$, one may still assume that the Fourier transform of uniform amplitude, $\tilde{U}(k)=\mathrm{const}$ for $\left\vert k\right\vert<k_{max}$, yields the most compact PSF. A quick calculation shows that this leads to a function proportional to $\sin(k_{max}x)/x$ which is not strictly non-negative, while strict non-negativity is a fundamentally important requirement for a physically valid PSF. Expanding on this idea, it is proposed that the most compact PSF is given by the \emph{square} of this function, $\left[\sin(k_{max}x/2)/x\right]^2$, which corresponds to the Fourier transform of the \emph{auto-convolution} of the re-scaled original uniform amplitude distribution, $\tilde{U}(2\mathbf{k})$, over half the support $\left\vert k\right\vert<k_{max}/2$, i.e. $\tilde{U}(2\mathbf{k}) \otimes \tilde{U}(2\mathbf{k})$. This auto-convolution does indeed have the original support $\left\vert k\right\vert<k_{max}$, and its Fourier transform is by definition strictly non-negative. Generalizing this idea to three dimensions, we arrive at the core hypothesis of our paper: For a given support $\mathcal{M}$, the most compact positive PSF is found as the Fourier transform of an OTF which is the auto-convolution of a uniform amplitude distribution. 

This problem is closely connected to Minkowski sums. A Minkowski sum of two sets of points $\mathcal{A}$ and $\mathcal{B}$ is defined by                

\begin{equation} \begin{split}
\mathcal{A} \oplus \mathcal{B} = \{ \mathbf{a}+\mathbf{b} \big\vert \mathbf{a} \in \mathcal{A} \wedge \mathbf{b} \in \mathcal{B} \}
\end{split} \end{equation}

\noindent If one has two functions $\tilde{f}(\mathbf{k})$ and $\tilde{g}(\mathbf{k})$ defined over finite supports $\mathcal{A}$ and $\mathcal{B}$, respectively, then their convolution $\tilde{f}(\mathbf{k})\otimes\tilde{g}(\mathbf{k})$ has the support $\mathcal{A} \oplus \mathcal{B}$. Thus, given the set of points $\mathcal{M}$ of an OTF's support, what we need to find is another set of points $\mathcal{M}^{1/2}$ so that 

\begin{equation} 
\begin{split}
\mathcal{M} = \mathcal{M}^{1/2} \oplus \mathcal{M}^{1/2}
\end{split} 
\label{eq:Minkowski}
\end{equation}

\noindent Knowing this set $\mathcal{M}^{1/2}$, the ideal OTF is found by an auto-convolution of a uniform amplitude distribution over $\mathcal{M}^{1/2}$, and the ideal PSF is the Fourier transform of this OTF. It is trivial to find $\mathcal{M}^{1/2}$ for a \emph{convex} set $\mathcal{M}$, which is $\mathcal{M}$ scaled down in extent by a factor of 2. For general sets $\mathcal{M}$, in particular non-convex or disjoint sets, there is no general algorithm known for finding $\mathcal{M}^{1/2}$. However, for many cases of practical relevance in microscopic imaging, one can find $\mathcal{M}^{1/2}$ by considering how the OTF is physically calculated. 

We will now show that a uniform phase and amplitude distribution across $\mathcal{M}^{1/2}$ gives rise to a maximum sharpness metric $U(0)$ as defined in eq.~\ref{eq:sharpnessmetric}. That is, given a Minkowski decomposition of the OTF support ${\cal M}$ in eq.~\ref{eq:Minkowski} we can construct the ideal OTF. A positive (and real) PSF can always be written as $U(\mathbf{r})=\left\vert E(\mathbf{r})\right\vert^{2}$, where the field distribution $E(\mathbf{r})$ can be written as a Fourier transform:

\begin{equation} 
\begin{split}
E(\mathbf{r}) = \int \frac{d\mathbf{k}}{(2\pi)^D} \tilde{E}(\mathbf{k}) \exp\left( i \mathbf{k}\cdot\mathbf{r}\right),
\end{split}
\label{eq:Ewidefield}
\end{equation}

\noindent where the support of $\tilde{E}(\mathbf{k})$ is $\mathcal{M}^{1/2}$. Without loss of generality we may write (for $\mathbf{k}\in \mathcal{M}^{1/2}$):

\begin{equation} 
\begin{split}
\tilde{E}(\mathbf{k}) = E_{0}\exp\left( i W_{c}(\mathbf{k}) \right),
\end{split}
\end{equation}

\noindent where $W_{c}(\mathbf{k})=W(\mathbf{k})+iV(\mathbf{k})$ is the complex aberration function, with real and imaginary parts $W(\mathbf{k})$ and $V(\mathbf{k})$ that represent the phase and amplitude aberration functions, respectively. The constant $E_0$ is fixed by the normalization condition:

\begin{equation} 
\begin{split}
1 &= \int d\mathbf{r} \left|E(\mathbf{r})\right|^{2} =\int \frac{d\mathbf{k}}{(2\pi)^D} \left|\tilde{E}(\mathbf{k})\right|^{2}\\
&= E_{0}^{2} \int_{\mathcal{M}^{1/2}} \frac{d\mathbf{k}}{(2\pi)^D} \left|\exp\left(iW_{c}(\mathbf{k}) \right)\right|^{2}.
\end{split}
\end{equation}

\noindent With the aid of this equation, $E_0$ can be solved and we find for the PSF peak value:

\begin{equation} 
\begin{split}
U(0) = \frac{1}{(2\pi)^D}\frac{ \left| \int_{\mathcal{M}^{1/2}} d\mathbf{k} \exp\left( iW_{c}(\mathbf{k}) \right)\right|^{2} }{\int_{\mathcal{M}^{1/2}} d\mathbf{k} \left|\exp\left(iW_{c}(\mathbf{k}) \right)\right|^{2}}.
\end{split}
\end{equation}

\noindent The PSF peak value is usually expressed in the dimensionless Strehl ratio:

\begin{equation} 
\begin{split}
S = \frac{U(0)}{A_p} = \frac{ \left| \langle  \exp\left( iW_{c}\right)\rangle \right|^{2} }{\langle \left|\exp\left(iW_{c} \right)\right|^{2}\rangle},
\end{split}
\end{equation}

\noindent where the \emph{pupil area} is defined as:

\begin{equation} 
\begin{split}
A_p = \int_{\mathcal{M}^{1/2}} \frac{d\mathbf{k}}{(2\pi)^D},
\end{split}
\end{equation}

\noindent and where the \emph{pupil average} of any function $F(\mathbf{k})$ is defined as:

\begin{equation} 
\begin{split}
\langle F \rangle = \frac{1}{A_p}\int_{\mathcal{M}^{1/2}} \frac{d\mathbf{k}}{(2\pi)^D} F(\mathbf{k}).
\end{split}
\end{equation}

\noindent Using $\left|\langle a|b\rangle\right|^{2}\leq\langle a|a\rangle\langle b|b\rangle$ for $a=\exp\left(iW_{c}\right)$ and $b=1$ we find that $S\leq 1$ and that the maximum is obtained in case $W_{c}(\mathbf{k})=0$, i.e. for zero phase and amplitude aberrations, leading to a uniform complex amplitude distribution over the pupil support $\mathcal{M}^{1/2}$. \added{This finalizes the proof, and enables a direct computation of the ideal OTF once the Minkowski decomposition of the OTF support is made.}

An interesting corollary of the current formalism is that to lowest order in the phase and amplitude aberration functions:

\begin{equation} 
\begin{split}
S = 1- W_{\mathrm{rms}}^{2}-V_{\mathrm{rms}}^{2},
\end{split}
\end{equation}

\noindent where the root mean square phase aberration is given by $W_{\mathrm{rms}}^{2}=\langle W^{2} \rangle - \langle W \rangle^{2}$ (and likewise for the root mean square amplitude aberration $V_{\mathrm{rms}}$). This suggests a generalization of Mar\'{e}chal's criterion for aberration tolerances to include amplitude aberrations as well. In fact, in the current form phase and amplitude aberrations can be treated on an equal footing. 

It appears that the decomposition of the OTF support ${\cal M}$ into the pupil ${\cal M}^{1/2}$ via the Minkowski sum is not unique. Consider for example the case of a circular support of a 2D-OTF. Then our procedure leads to a circular pupil scaled down twofold in size compared to the OTF support. An annular pupil, however, is an equally valid solution ${\cal M'}^{1/2}$ to the Minkowski sum decomposition. Such pupil shapes can be incorporated into the current formalism because these sets ${\cal M'}^{1/2}$ are subsets of the full set ${\cal M}^{1/2}$, i.e. ${\cal M'}^{1/2}\subset{\cal M}^{1/2}$ for any solution ${\cal M'}^{1/2}$ to the Minkowski sum decomposition. This can be taken into account by an amplitude aberration function $V(\mathbf{k})=0$ for $\mathbf{k}\in{\cal M'}^{1/2}$ and $V(\mathbf{k})=\infty$ for $\mathbf{k}\notin{\cal M'}^{1/2}$. Then the Strehl-ratio can be computed as

\begin{equation} 
\begin{split}
S = \frac{1}{A_{p}}\int_{\mathcal{M'}^{1/2}} \frac{d\mathbf{k}}{(2\pi)^D},
\end{split}
\end{equation}

\noindent which satisfies $S\leq 1$, i.e. the overall performance, quantified by the summed OTF over the support ${\cal M}$, is always worse. This fits with the well-known behavior that annular pupils can give rise to a narrower central peak of the PSF, but always at the expense of enhanced side lobes \cite{toraldo1952annularpupils,cox82annularpupils,rogers2012sol}. We mention that for a small range of spatial frequencies the use of annular pupils possibly gives rise to a higher OTF than out ideal OTF, but for the average OTF over all spatial frequencies this does not appear to be the case.

In what follows, we will demonstrate the application of our concept of an optimal PSF/OTF on four examples: a wide-field microscope, an image scanning microscope, a 4pi-microscope, and a structured illumination microscope.

\section{Wide-field Microscope} 

\begin{figure}[hbt]
\centering
\includegraphics[width=\linewidth]{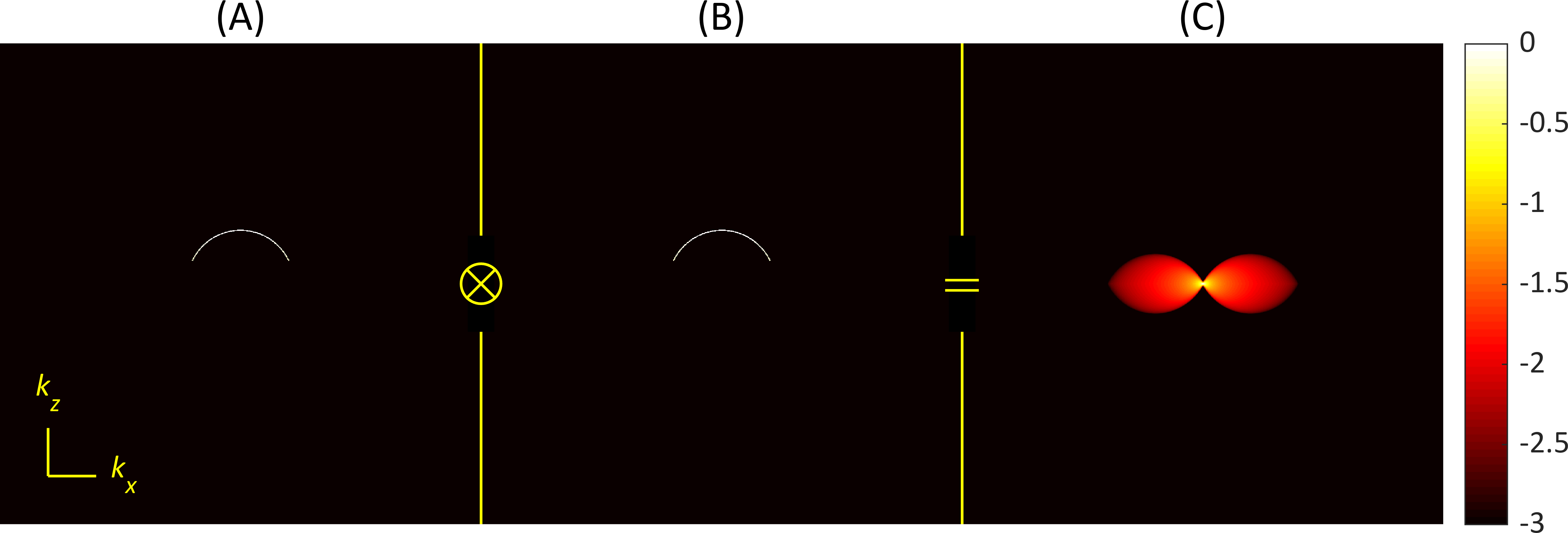}
\caption{OTF of a wide-field microscope. (A,B) $(k_x,k_z)$-cross section of the axi-symmetric Fourier transform $\tilde{E}(\mathbf{k})$ of the electric field distribution, eq.~\eqref{eq:Ewidefield}. Optical axis is along $k_z$. Extent along the $k_x$-direction is $\pm k\mathrm{NA}/n$, and along the $k_z$-direction $k[1-\sqrt{1-(\mathrm{NA}/n)^2}]$. Numerical aperture NA of the microscope's objective was set to 1.2, refractive index $n$ of objective's immersion medium (sample space) to 1.33 (water immersion), and image magnification $M$ to 60. (C) OTF of wide-field microscope as convolution of (A) with (B), i.e. the auto-convolution of $\tilde{E}(\mathbf{k})$ with an extent double as large in all directions as that of the Fourier transform in (A,B). All three distributions in (A-C) are normalized by their maximum value.}
\label{fig:OTFwidefield}
\end{figure}

As a first example, we consider a perfectly imaging, aberration-free wide field microscope equipped with an objective of numerical aperture NA. Here and it the next sections, we will work in the scalar approximation of optical imaging, i.e. neglecting the vector character of the electric-magnetic field -- discussion and justification of our results in the light of a full vector theory of electro-magnetic radiation will be given in Section ``High numerical aperture and polarization effects'' at the end of this paper. Thus, following the scalar theory of imaging through a high-aperture optics, the electric field distribution of the image of a point emitter is given by eq.~\eqref{eq:Ewidefield} where the finite \emph{two-dimensional} integration domain is an axi-symmetric cap of the sphere $\left\vert \mathbf{k}\right\vert=n k_0$ defined by $\arctan(k_\perp/k)\leq\theta_{max}$, with $k_0$ being the length of the emission light's wave vector in vacuum, $n$ the refractive index of the objective's immersion medium (sample space), $k_\perp$ the modulus of the transverse component of $\mathbf{k}$ perpendicular to the optical axis, and $\theta_{max}$ the maximum half angle of light collection of the microscope's objective. This angle is connected with the objective's numerical aperture by $\mathrm{NA}=n \sin\theta_{max}$. A cross section of the integration domain is shown in Fig.~\ref{fig:OTFwidefield}(A). It is important to note that the above electric field distribution is given as a function of the coordinate $\mathbf{r}$ in \emph{sample space} -- the corresponding transverse position in image space is the transverse component of $\mathbf{r}$ multiplied by magnification $M$. 

For an isotropically emitting point source, energy conservation implies that the amplitude function $\tilde{E}(\mathbf{k})$ is equal to $\sqrt{\cos\theta/\cos\theta'}$, where $\theta=\arctan(k_\perp/k_\parallel)$ with $k_\parallel$ being the component of the wave vector along the optical axis, and the angle $\theta'$ in image space is connected to $\theta$ via Abbe's sine condition, $M n' \sin\theta' = n \sin\theta$, where $M$ is the image magnification, and $n'$ the refractive index of image space (typically $n'=1$) \cite{wolf1959electromagnetic,richards1959electromagnetic}. 

\begin{figure}[hbt]
\centering
\includegraphics[width=\linewidth]{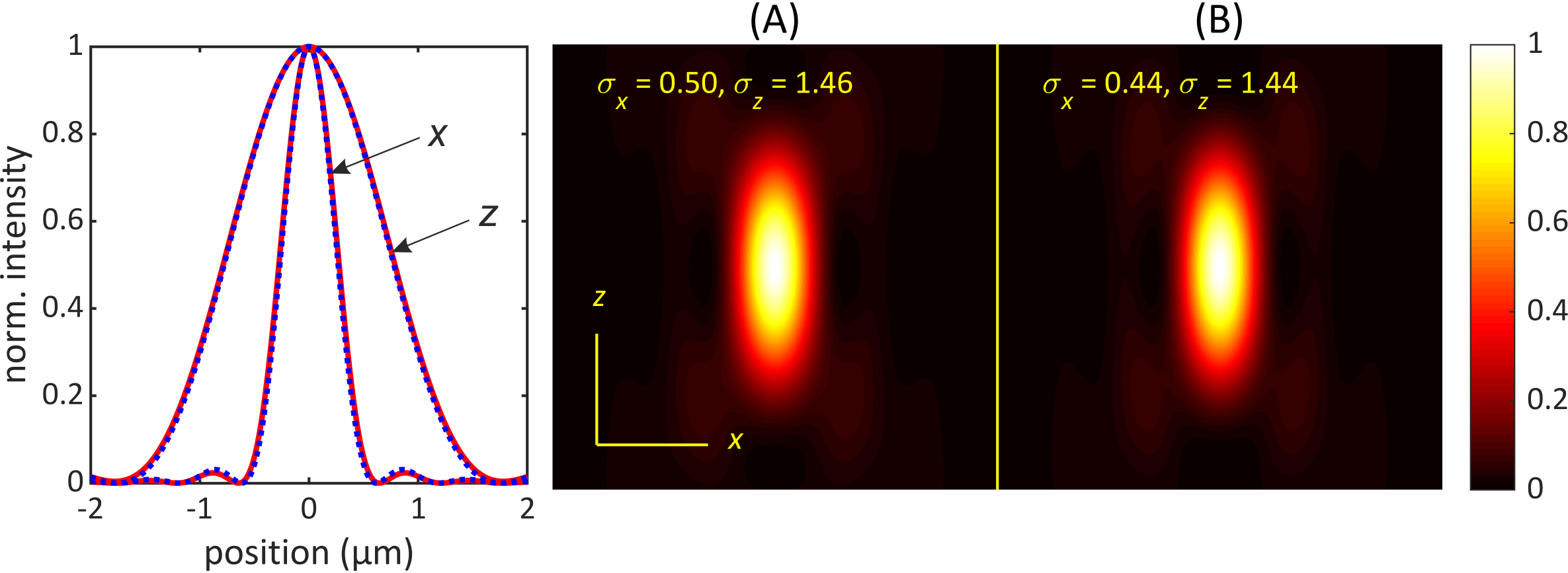}
\caption{Comparison of actual and ideal PSFs of a wide-field microscope. (A) PSF obtained via a Fourier transform of the OTF shown in Fig.~\ref{fig:OTFwidefield}(C). Length of shown yellow coordinate vectors is one wavelength. (B) Ideal PSF obtained from the OTF calculated as the auto-convolution of two \emph{uniform} amplitude distributions with the same frequency support as those shown in Fig.~\ref{fig:OTFwidefield}(A,B). At the top of both panels are indicated the square root variances (in units of wavelength) of the distributions when \added{fitting their lateral and axial projections with Gaussian functions. The curves on the left show cross-sections along the $x$- and $z$-directions for both panel (A) (solid lines) and panel (B) (dotted lines).}}
\label{fig:PSFwidefield}
\end{figure}

The PSF is the absolute square of the electric field distribution, $U(\mathbf{r}) = \left\vert \mathbf{E}(\mathbf{r}) \right\vert^2$, which means that the OTF is the three-dimensional auto-convolution of $\tilde{E}(\mathbf{k})$, as shown in Fig.~\ref{fig:OTFwidefield}. 
For axi-symmetric functions such as $\tilde{E}(\mathbf{k})$ considered here, such auto-convolutions can be computed using the theory of Ref.~\cite{sheppard1994}. We follow the numerically efficient and more generally applicable method using Hankel transforms, see SI of Ref.~\cite{schulz2013resolution}.
The resulting PSF, i.e. Fourier transform of the OTF shown in Fig.~\ref{fig:OTFwidefield}(C), is presented in Fig.~\ref{fig:PSFwidefield}(A).

Following our hypothesis, the most optimal OTF is obtained by an auto-convolution of a uniform amplitude distribution yielding the same support as the original OTF. As already mentioned, this reduces to finding a set of points $\mathcal{M}^{1/2}$ so that its Minkowski sum with itself yields the support of the OTF in Fig.~\ref{fig:OTFwidefield}(C). For such non-convex sets, no general solution for finding $\mathcal{M}^{1/2}$ is known, but in the present case, it is obviously known from the very physics of how the OTF is calculated, and is the support of the Fourier transform of the electric field as shown in Fig.~\ref{fig:OTFwidefield}(A,B). Thus, the ideal OTF is found by replacing in \eqref{eq:Ewidefield} the $\mathbf{k}$-dependent amplitude by a constant one, $\tilde{E}(\mathbf{k})=\mathrm{const}$, and then auto-convolving the result. We do not show the resulting OTF, which looks very similar to the original one, but present the result for the corresponding PSF in Fig.~\ref{fig:PSFwidefield}(B). One finds indeed a slight improvement in spatial resolution (quantified by the square root of the variance values of three-dimensional Gaussian distributions fitted to the PSFs). However, the improvement in lateral resolution is only ca. 10\%, and in axial resolution it is negligible. Thus, a perfectly adjusted wide-field microscope operates close to the absolute optimum in the sense that it transmits the spatial frequencies of the sample in a close-to-optimum manner. 

\section{Image Scanning Microscopy} 

\begin{figure}[hbt]
\centering
\includegraphics[width=\linewidth]{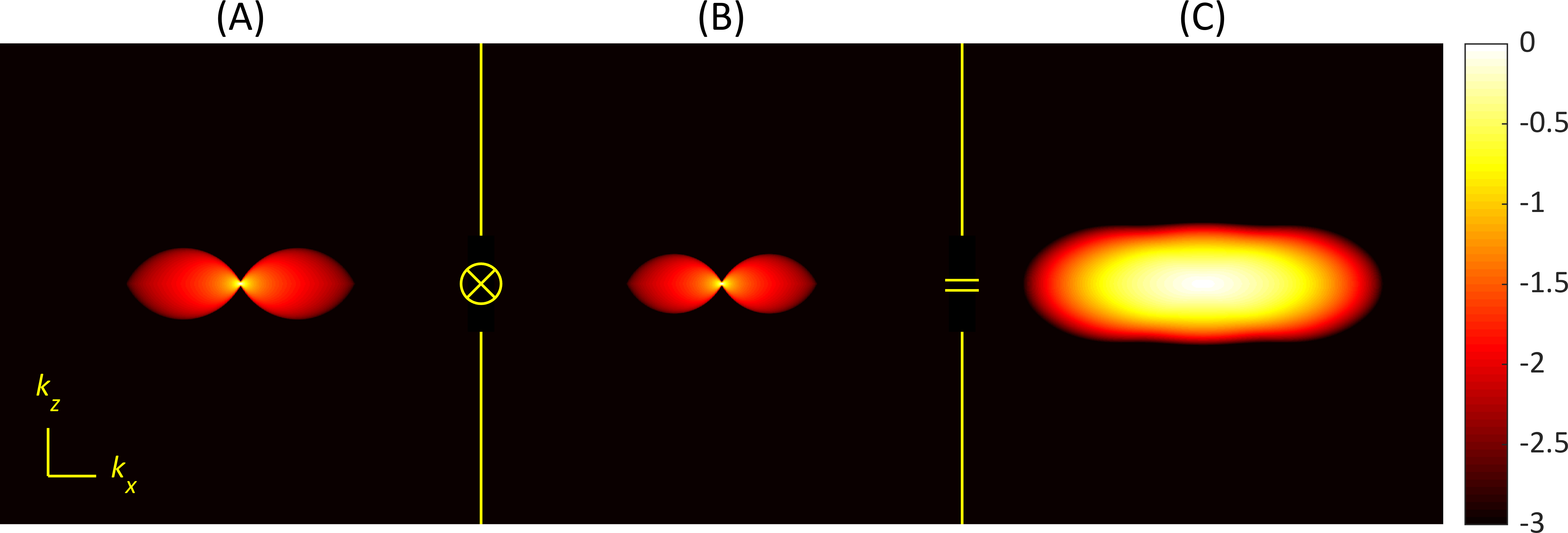}
\caption{OTF of ISM. (A) Auto-convolution of the Fourier transform of the focused excitation electric field, eq.~\eqref{eq:Ewidefield}, with $\tilde{E}(\mathbf{k})=\sqrt{\cos\theta}$, and for a modulus of the wave vector which is assumed to be 6/5 that of the fluorescence emission light. (B) OTF of light detection efficiency distribution for light detection with an infinitely small pinhole. All other microscope parameters were again set to the same values as for the wide-field microscopy example, i.e. $\mathrm{NA}=1.2$, $n=1.33$, and $M=60$. (C) OTF of ISM obtained by the convolution of (A) with (B).}
\label{fig:OTFISM}
\end{figure}

Next, we consider an Image Scanning Microscope (ISM), that scans the sample with a diffraction-limited focus, and records at each scan-position an image of the excited fluorescence. From the resulting 3+2-dimensional data set (three dimensions for scan positions, two dimensions for the images taken at each scan position), a final image is calculated that is equivalent to an image recorded with a conventional laser-scanning confocal microscope having an infinitely small confocal aperture. Thus, the PSF of the final ISM image is given by the product of the excitation intensity distribution and the light collection efficiency of detecting light through an infinitely small confocal aperture. The Fourier transform of the former is given by the auto-convolution of the Fourier representation of the electric field in sample space when focusing a perfectly plane wave through the objective. This Fourier representation is given again by \eqref{eq:Ewidefield}, but now with amplitude function $\tilde{E}(\mathbf{k})=\sqrt{\cos\theta}$, and at the wavelength of the excitation light (which is smaller than that of the fluorescence emission, resulting in a larger modulus $k$ of the wave vector). This auto-convolution of this Fourier transform is shown in Fig.~\ref{fig:OTFISM}(A). The Fourier transform of the light collection efficiency for an infinitely small pinhole is the same as the OTF of a wide field microscope as shown in Fig.~\ref{fig:OTFwidefield}(C), and is reproduced in Fig.~\ref{fig:OTFISM}(B). Finally, the result for the OTF of ISM is shown in Fig.~\ref{fig:OTFISM}(C), and the corresponding PSF in Fig.~\ref{fig:PSFISM}(A).

\begin{figure}[hbt]
\centering
\includegraphics[width=\linewidth]{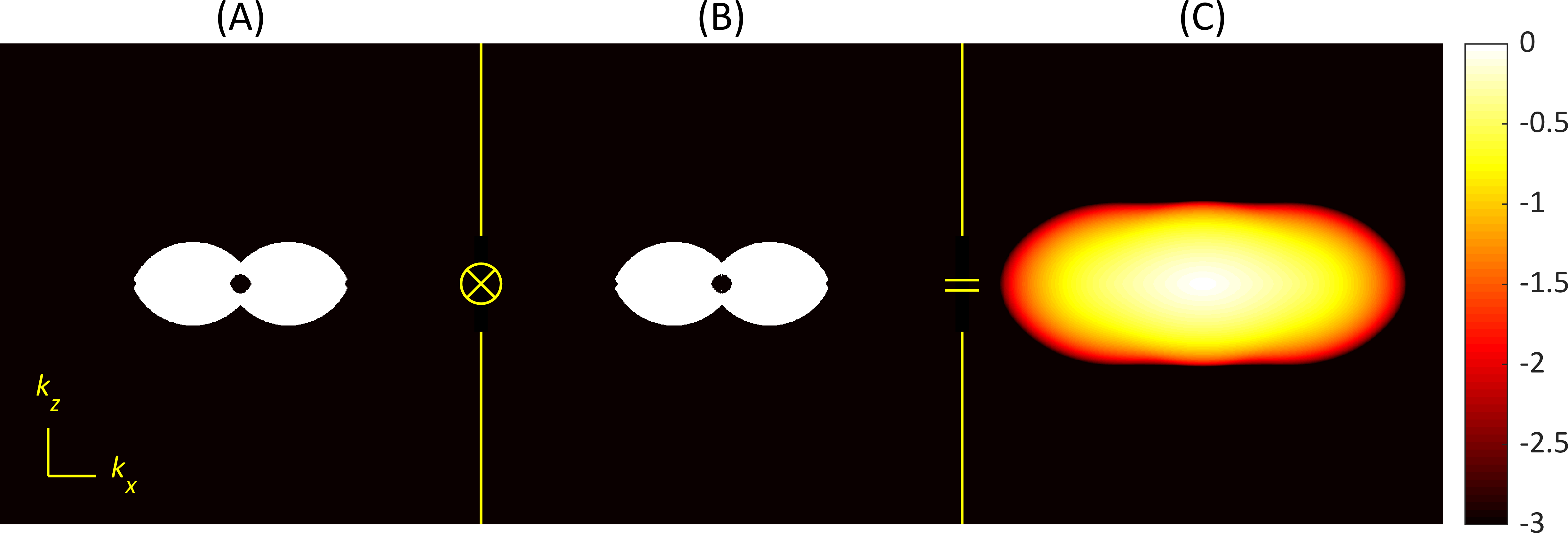}
\caption{Ideal OTF of ISM. (A,B) Uniform amplitude distributions over the frequency support of the convolution of the Fourier transforms of excitation and detection electric fields. Please not the hole of missing frequencies in the middle which is due to the different wavelengths (and thus lengths of the wave vectors) of excitation and emission. (C) Ideal OTF of ISM as obtained by the auto-convolution of (A,B).}
\label{fig:IdealOTFISM}
\end{figure}

\begin{figure}[hbt]
\centering
\includegraphics[width=\linewidth]{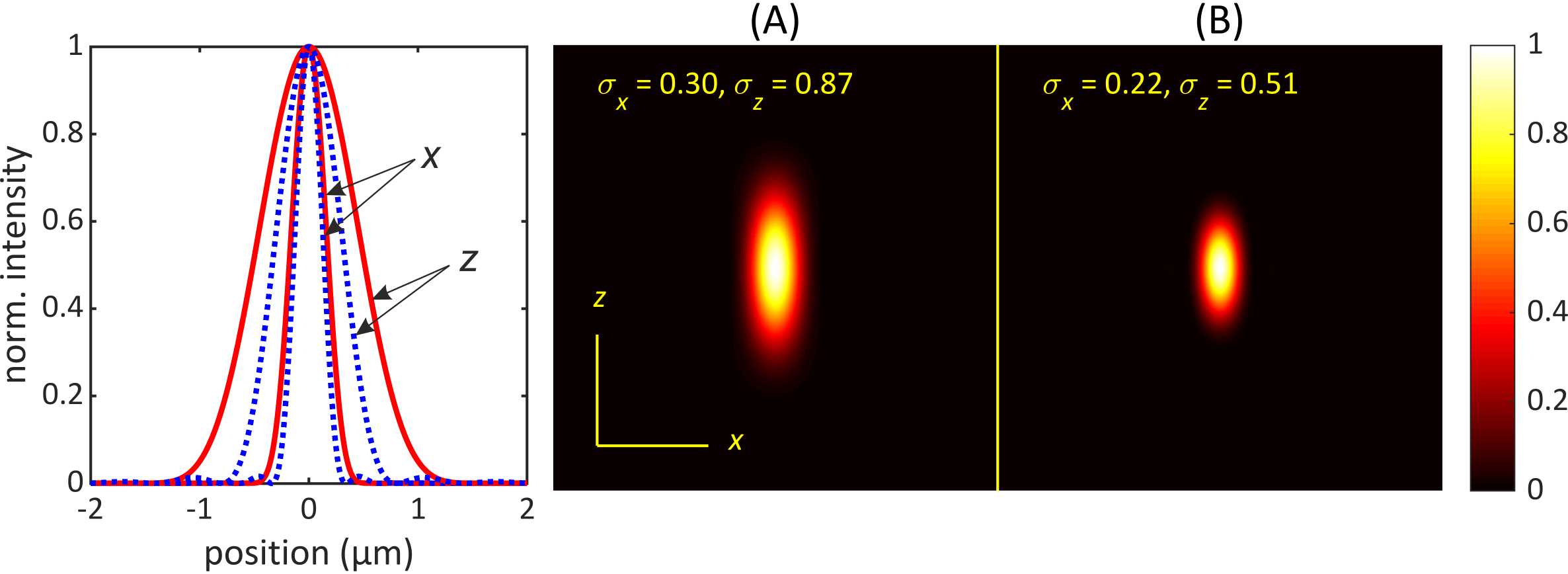}
\caption{(A) PSF of ISM as obtained via the Fourier transform of the OTF show in Fig.~\ref{fig:OTFISM}(C). (B) Ideal PSF of ISM as obtained by a Fourier transform of the ideal OTF shown in Fig.~\ref{fig:IdealOTFISM}(C). \added{The curves on the left show cross-sections along the $x$- and $z$-directions for both panel (A) (solid lines) and panel (B) (dotted lines).}}
\label{fig:PSFISM}
\end{figure}

Again, following our core hypothesis, the ideal OTF is obtained as an auto-convolution of a uniform amplitude distribution yielding the same support as that of the OTF shown in Fig.~\ref{fig:OTFISM}(C). For finding the corresponding $\mathcal{M}^{1/2}$, we can use the associative property of Minkowski addition: For example, for two sets $\mathcal{A}$ and $\mathcal{B}$ we have $(\mathcal{A}\oplus\mathcal{A})\oplus(\mathcal{B}\oplus\mathcal{B})=(\mathcal{A}\oplus\mathcal{B})\oplus(\mathcal{A}\oplus\mathcal{B})$. Thus, in our case of ISM, the sought-after $\mathcal{M}^{1/2}$ is the support of the convolution of the Fourier transform of the excitation electric field with the Fourier transform of the detection electric field. This is shown in Fig.~\ref{fig:IdealOTFISM}(A,B). Thus, the ideal OTF of ISM is found as the auto-convolution of a uniform amplitude distribution over this support and is shown in Fig.~\ref{fig:IdealOTFISM}(C). The resulting PSF is presented in Fig.~\ref{fig:PSFISM}(B). As can be seen, for ISM, the optimal amplitude distribution yields a significantly smaller PSF, and thus an appropriate deconvolution of ISM images using this PSF can indeed significantly improve image resolution and contrast (brightness per image area) \cite{muller2010image}.  

\section{4pi Microscopy} 

\begin{figure}[hbt]
\centering
\includegraphics[width=\linewidth]{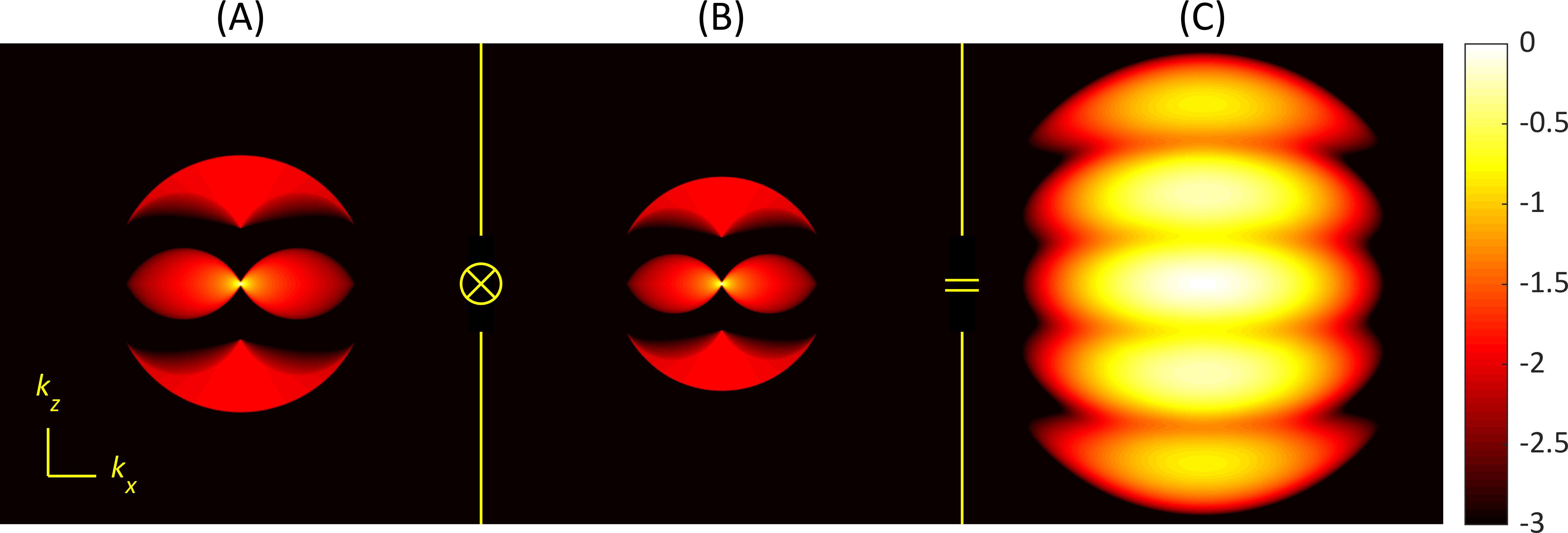}
\caption{OTF of a type-C 4pi microscope. (A) Fourier transform of the excitation intensity distribution. Again, excitation wavelength is set to 5/6 of fluorescence emission wavelength. (B) Fourier transform of the light detection efficiency distribution. All microscope parameters were again set to the same values as for the wide-field microscopy example, i.e. $\mathrm{NA}=1.2$, $n=1.33$, and $M=60$. (C) OTF of a type-C 4pi microscope obtained by the convolution of (A) with (B).}
\label{fig:OTF4pi}
\end{figure}

As a third example, we consider the case of a so-called type-C 4pi microscope \cite{hell1992properties,Schrader_Kozubek_Hell_Wilson_1997}. In such a microscope, the sample is illuminated from both sides through two identical objectives that constructively focus the light of two mutually coherent laser beams, and the excited fluorescence emission is detected through the same two objectives by constructive interference, as is done also in 3D interferometric photoactivated localization fluorescence microscopy (iPALM) \cite{shtengel2009interferometric} or isotropic stimulated emission depletion (isoSTED) microscopy \cite{schmidt2008spherical}. The Fourier representation of the excitation electric field is again given by eq.~\eqref{eq:Ewidefield}, but now with an integration domain $\Omega$ consisting of \emph{two} segments of the sphere $\left\vert \mathbf{k}\right\vert=n k_0$ defined by $\left\vert\arccos(k_\parallel/k_\perp)\right\vert\leq\theta_{max}$. The resulting distribution looks very similar to the one shown in Fig.~\ref{fig:OTFwidefield}(A), but with a a second spherical segment that is a mirror image of the one shown reflected on the horizontal axis, and both having a larger radius (due to the excitation wavelength being shorter than the emission wavelength). The auto-convolution of such a distribution and thus the Fourier transform of the excitation intensity distribution is presented in Fig.~\ref{fig:OTF4pi}(A). The Fourier transform of the light collection efficiency distribution is obtained analogously and is shown in Fig.~\ref{fig:OTF4pi}(B). Thus, the OTF of a type-C 4pi microscope is finally obtained by the convolution of both Fourier transforms and shown in Fig.~\ref{fig:OTF4pi}(C). The corresponding PSF is seen in Fig.~\ref{fig:PSF4pi}(A). 

\begin{figure}[hbt]
\centering
\includegraphics[width=\linewidth]{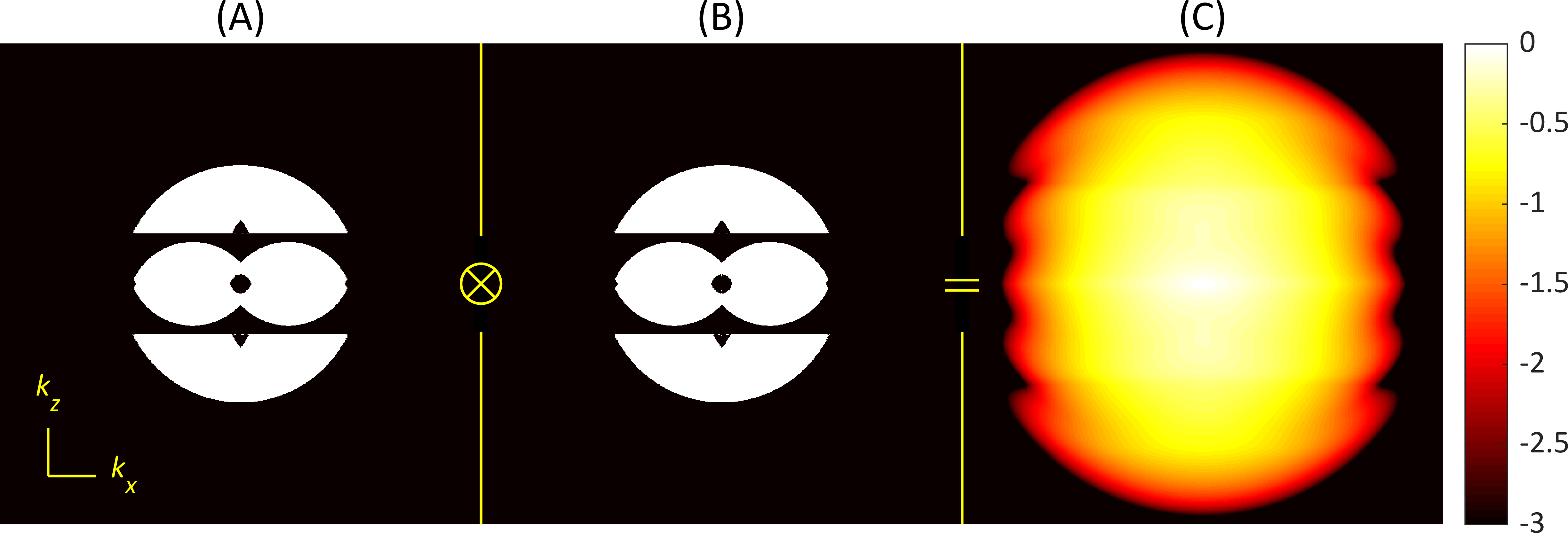}
\caption{Ideal OTF of a type-C 4pi microscope. (A,B) Uniform amplitude distributions over the frequency support of the convolution of the Fourier transforms of excitation and detection electric fields. (C) Ideal OTF of a type-C 4pi microscope as obtained by the auto-convolution of (A,B).}
\label{fig:IdealOTF4pi}
\end{figure}

\begin{figure}[hbt]
\centering
\includegraphics[width=\linewidth]{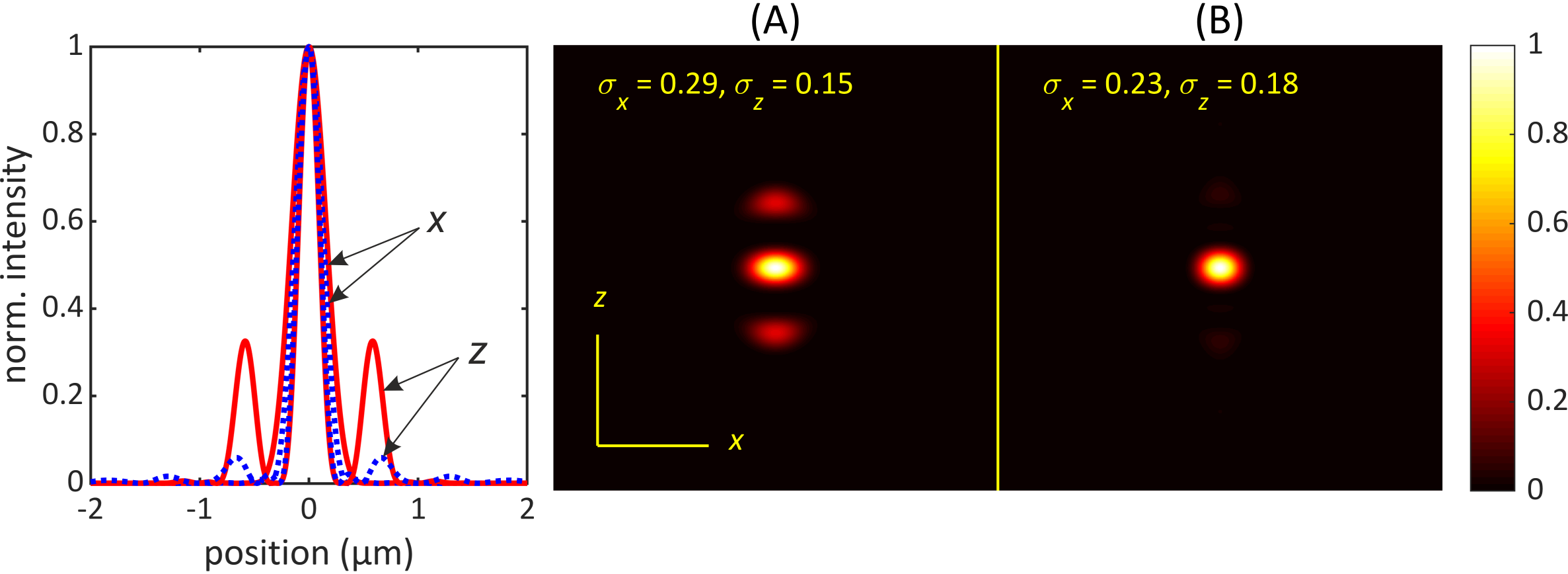}
\caption{(A) PSF of a type-C 4pi microscope, being the Fourier transform of the OTF seen in Fig.~\ref{fig:OTF4pi}(C). (B) Ideal PSF of a type-C 4pi microscope calculated as the Fourier transform of the OTF seen in Fig.~\ref{fig:IdealOTF4pi}(C). \added{The given values of square root variances, $\sigma_{x,z}$, refer to fits of the lateral and axial projections of the central peaks of the PSFs with Gaussian functions. The curves on the left show cross-sections along the $x$- and $z$-directions for both panel (A) (solid lines) and panel (B) (dotted lines).}}
\label{fig:PSF4pi}
\end{figure}

Using the same considerations as in the previous example, the set $\mathcal{M}^{1/2}$ is obtained as the support of the convolution of the Fourier electric field representations for excitation and detection, and is shown in Fig.~\ref{fig:IdealOTF4pi}(A,B). The auto-convolution of a uniform amplitude distribution over this support then yields the ideal OTF of a type-4 4pi microscope and is shown in Fig.~\ref{fig:IdealOTF4pi}(C). Its Fourier transform then gives the ideal PSF as presented in Fig.~\ref{fig:PSF4pi}(B). As can be seen, this PSF is in size not much different from that of the central peak of the original one, but now the axial side peaks are significantly suppressed. Using this ideal PSF for image deconvolution promises an efficient suppression of ghost images due to the axial side peaks in the original PSF, which is a serious issue in 4pi microscopy.

\section{Structured Illumination Microscopy}

\begin{figure}[hbt]
\centering
\includegraphics[width=\linewidth]{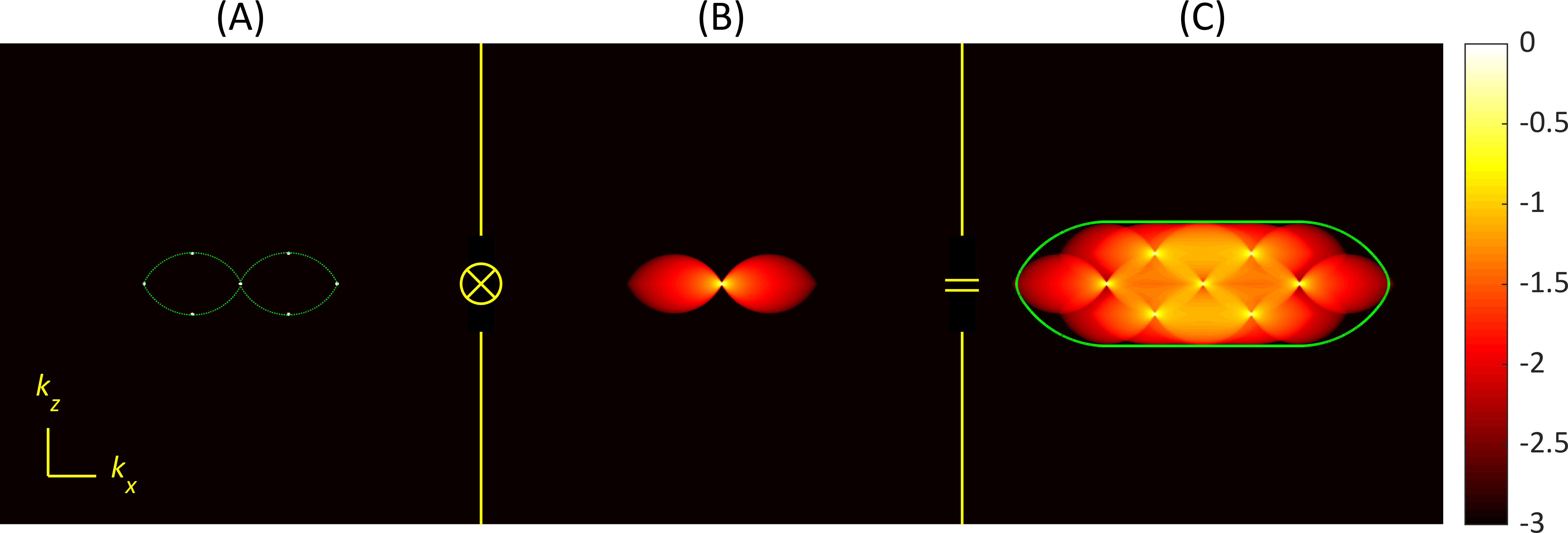}
\caption{Comparison of spectral support of the OTFs for 3D SIM and for ISM. For the sake of simplicity, the excitation and detection wavelengths are set equal. (A) OTF of excitation intensity distribution of 3D SIM. Shown are the 7 peaks at the center $\left\{0,0,0\right\}$ and positions $\mathbf{k} = n k_0 \left\{\pm 2\sin\theta_{max},0,0\right\}$ and $\mathbf{k} = n k_0 \left\{\pm \sin\theta_{max},0,\pm 2(1-\cos\theta_{max})\right\}$. The green dotted line shows the frequency support of the corresponding OTF for ISM excitation, which is identical to the OTF shown in the middle panel. (B) OTF of wide-field detection. (C) Superposition of (B) shifted to all peak positions shown in (A) plus all additional peak positions obtained by rotating the peaks around the vertical (optical) axis in steps of 60$^\circ$. The result visualizes the coverage of the frequency space by 3D SIM. Green line shows the limit of the frequency support for the ISM OTF, obtained by auto-convolution of (B).}
\label{fig:IdealOTFSIM}
\end{figure}

As a last example we consider a three-dimensional structured illumination microscope (3D SIM) \cite{Gustafsson_Shao_Carlton_Wang_Golubovskaya_Cande_Agard_Sedat_2008,Kraus_Miron_Demmerle_Chitiashvili_Budco_Alle_Matsuda_Leonhardt_Schermelleh_Markaki_2017,heintzmann2017simreview}. This example is different from the previous ones because the OTF and PSF now, as will be seen, are no longer axially symmetric. In a 3D SIM microscope, illumination is done by the superposition of three plane waves, one traveling along the optical axis and two with propagation angle $\theta=\pm \arcsin(\beta\mathrm{NA}/n)$ with respect to the optical axis while sharing a common incidence plane. Here $\beta$ is a numerical coefficient that is chosen to be close to 1. For the sake of simplicity we will set it to the best possible value $\beta=1$ in the following. This generates a 3D periodic excitation intensity pattern in sample space. The Fourier transform of this intensity distributions consists of seven delta-function peaks in the plane of incidence of the inclined excitation plane waves and is shown in Fig.~\ref{fig:IdealOTFSIM}(A). The OTF of detection of the 3D SIM is the same as that of a wide-field microscope, Fig.~\ref{fig:OTFwidefield}(C), and is shown, for the sake of completeness, again in Fig.~\ref{fig:IdealOTFSIM}(B). By taking several images for different relative positions and orientations (usually by rotating around the optical axis in angular steps of 60$^\circ$) of this excitation intensity pattern with respect to a sample, and applying a sophisticated image reconstruction algorithm, one finally obtains an image with doubled lateral resolution (as compared to a wide-field microscope) and an axial resolution and sectioning comparable to a confocal microscope.

It is difficult to define a generic OTF for 3D SIM, because the OTF (and PSF) will depend on the specific details of the image reconstruction algorithm used, such as the different spatial frequency filtering steps. Whatever the image reconstruction algorithm is, the resulting OTF will have the same support as the linear superposition of copies of the distribution shown in Fig.~\ref{fig:IdealOTFSIM}(B), shifted to the peak positions of Fig.~\ref{fig:IdealOTFSIM}(A). This is visualized in Fig.~\ref{fig:IdealOTFSIM}(C) showing a cross section of such a superposition obtained by shifting the OTF of Fig.~\ref{fig:IdealOTFSIM}(C) to all 19 peak positions in 3D, 7 as shown in Fig.~\ref{fig:IdealOTFSIM}(A), and 12 additional from rotations of the non-axial peaks around the vertical axis by 60$^\circ$ and 120$^\circ$. This cross-section visualizes the frequency range covered by SIM.
A direct application of our procedure, i.e. finding the effective pupil support $\mathcal{M}^{1/2}$ given the OTF support $\mathcal{M}$, appears to be too involved in view of the complex rotationally asymmetric shape of the OTF support for 3D SIM. A good approximation, however, can be found by comparison to the case of ISM.
The green line in Fig.~\ref{fig:IdealOTFSIM}(C) delimits the frequency support of the OTF of an ideal ISM (for the situation of no Stokes shift) as obtained through an auto-convolution of the distribution of Fig.~\ref{fig:IdealOTFSIM}(B). As can be seen, SIM covers nearly the same frequency space as ISM, and thus a fairly good approximation of the optimal OTF of 3D SIM is given by that of the corresponding ISM (same excitation and emission wavelengths). Thus, the ideal OTF (and PSF) of 3D SIM is expected to be very similar and well approximated by that of ISM as derived in the section about ISM above. 

\section{High numerical aperture and polarization effects} 

Up to now, all above considerations were based on a scalar approximation of image formation, thus neglecting the vector character of the electromagnetic field and corresponding polarization effects in fluorescence excitation and imaging, which become especially important for objectives with high numerical aperture. However, as will be explained in the following, our considerations and definition of the ideal PSF and OTF will be fully valid also for the general vectorial case. As an example, let us consider image formation in a conventional wide-field microscope. Following the wave-vector theory of Richardson and Wolf \cite{richards1959electromagnetic,wolf1959electromagnetic}, a point dipole emitter with dipole amplitude vector $\mathbf{d}$ positioned in the focal plane and on the optical axis generates an electric field distribution in the image plane of the microscope that is proportional to the following plane-wave superposition:

\begin{equation} 
\begin{split}
\mathbf{E}(\mathbf{r}) = &\int_0^{2\pi} d\psi \int_0^{\eta_{max}} d\eta \sin\eta \sqrt{\frac{\cos\eta}{\cos\eta'}} \cdot \\
&\Bigg.\left[ \hat{\mathbf{e}}_p \left(\hat{\mathbf{e}}'_p\cdot\mathbf{d}\right) + \hat{\mathbf{e}}_s \left(\hat{\mathbf{e}}_s\cdot\mathbf{d}\right)\right] e^{i \mathbf{k}\cdot\mathbf{r}}
\end{split} 
\label{eq:Evector}
\end{equation}

\noindent where $\eta$ and $\psi$ are azimuthal and polar integration angles, the relation between angles $\eta$ and $\eta'$ is given by Abbe's sine condition, $M \sin\eta = n\sin\eta'$, with $M$ being the image magnification and $n$ the refractive index of the objective's immersion medium (ideally equal to that of the sample), $\mathbf{k}=-k_0\left\{ \sin\eta \cos\psi, \sin\eta \sin\psi, \cos\eta  \right\}$ is the wave vector along the propagation direction of the plane waves, with $k_0=2\pi/\lambda$ its length at wavelength $\lambda$, and $\hat{\mathbf{e}}_p=\left\{ \cos\eta \cos\psi,\cos\eta \sin\psi, -\sin\eta \right\}$, $\hat{\mathbf{e}}'_p=\left\{ \cos\eta' \cos\psi, \cos\eta' \sin\psi, \sin\eta' \right\}$ and $\hat{\mathbf{e}}_s=\left\{ -\sin\psi, \cos\psi, 0 \right\}$ are azimuthal and polar polarization unit vectors. The maximum integration value $\eta_{max}$ is defined by the numerical aperture NA of the objective via $\eta_{max}=\arcsin\left(\text{NA}/M\right)$. The factor $\sqrt{\cos\eta/\cos\eta'}$ assures energy conservation during imaging. The corresponding magnetic field distribution is given by a similar plane wave superposition

\begin{equation} 
\begin{split}
\mathbf{B}(\mathbf{r}) = &\frac{1}{c}\int_0^{2\pi} d\psi \int_0^{\eta_{max}} d\eta \sin\eta \sqrt{\frac{\cos\eta}{\cos\eta'}} \cdot \\
&\Bigg.\left[ -\hat{\mathbf{e}}_s \left(\hat{\mathbf{e}}'_p\cdot\mathbf{d}\right) + \hat{\mathbf{e}}_p \left(\hat{\mathbf{e}}_s\cdot\mathbf{d}\right)\right] e^{i \mathbf{k}\cdot\mathbf{r}}
\end{split} 
\label{eq:Bvector}
\end{equation}

\noindent and the final intensity distribution in the image plane is given by the time-averaged Poynting vector component along the optical axis, which is

\begin{equation} 
\begin{split}
U_\mathbf{d}(\mathbf{r}) = -\frac{c}{8\pi}\Re\left[\hat{\mathbf{e}}_z\cdot(\mathbf{E}\times\mathbf{B})\right]
\end{split} 
\end{equation}

\noindent with $\hat{\mathbf{e}}_z$ being a unit vector along the optical axis ($z$-axis). Please note that all equations were written in such a coordinate system where the light from the dipole to the image plane propagates along the negative $z$-direction. For an isotropic distribution of emitters, the last expression has still to be averaged over all dipole orientations which then yields the final PSF of the system as 

\begin{align}
U(\mathbf{r}) = \langle U_\mathbf{d}(\mathbf{r}) \rangle_\mathbf{d} = -\frac{c}{8\pi}\langle\Re\left[\mathbf{e}_z\cdot(\mathbf{E}\times\mathbf{B})\right]\rangle_\mathbf{d}
\label{eq:Uvector}
\end{align}

\noindent where $\langle\cdot\rangle_\mathbf{d}$ denotes averaging over all orientations of $\mathbf{d}$.

Although this expression for the PSF looks now much more complicated than what we have seen for all PSFs in the scalar approximation of the previous sections, the core difference is the presence of the vectorial prefactors in front of the exponents $\exp(i\mathbf{k}\cdot\mathbf{r})$ in the above integrals, so that the integrals return vector functions instead of scalar functions. However, when comparing these integrals with the scalar approximation of eq.~\ref{eq:Ewidefield}, one can see that \emph{the frequency support of the Fourier transforms of all these functions is identical}! Furthermore, the PSF calculated with the vectorial representations of the electric and magnetic field is given by the real part of a cross product of both fields, which corresponds in Fourier space to a sum of several scalar convolutions. As a result, also the frequency support of this Fourier-transformed PSF is identical to that of the OTF calculated in the scalar approximation. The vector character of the electric and magnetic field makes the calculations of the PSF and OTF much more complicated than in the scalar approximation, but it does not at all change the frequency support of the final OTF. This support remains the same, irrespective of any alterations to the amplitude, phase, or polarization in the pupil or the light path towards the detector.

Thus, all our previous derivations of ideal PSFs and OTFs remain fully intact, and one can use them with same validity for optimally deconvolving images that are obtained with high-NA objectives where vector effects can no longer be neglected. Even more, in the case of a wide-field microscope, the complex vectorial structure of the electric and magnetic fields will lead to a PSF/OTF that is sub-optimal (because it will inadvertently lead to a non-uniform amplitude distribution of the resulting OTF in Fourier space), but which can now be improved by deconvolving an image using the ideal OTF as derived in the scalar approximation. It remains an open question how close the ideal OTF can be approximated in a high-NA wide-field microscope by engineering the phase, amplitude, and polarization in the pupil plane. This would amount to maximizing the $z$-component of the Poynting vector in focus, given the total power flowing through the microscope's pupil towards the detector. An approach similar to ref.~\cite{urbach2008maxfield}, where the electric field magnitude in focus is optimized, could be a good starting point.   

\section{Application in linear deconvolution}
The concept of an ideal OTF is important for often applied linear deconvolution methods. Let us assume, the actual experimental PSF of a given microscope is $U_{exp}(\mathbf{r})$ which can take into account all kinds of imperfections such as optical aberrations, angle-dependent transmission efficiency of the objective etc. These imperfections will affect the measured image $I_{exp}(\mathbf{r})$ (blurring, reducing contrast, etc.). If one knows the ideal OTF $\tilde{U}(\mathbf{k})$ over the same frequency support as that of $\tilde{U}_{exp}(\mathbf{k})$, one can calculate a rectified Fourier-transformed image as
\begin{equation} \begin{split}
\tilde{I}(\mathbf{k}) = \frac{\tilde{U}(\mathbf{k})}{\tilde{U}_{exp}(\mathbf{k})} \tilde{I}_{exp}(\mathbf{k}).
\label{eq:filtering}
\end{split} \end{equation}
Just dividing by $\tilde{U}_{exp}(\mathbf{k})$ seemingly results in a better, more sharp, outcome of the deconvolution. In that case, however, the effective OTF will be essentially flat across the OTF support ${\cal M}$, which will violate the PSF positivity constraint. The key thesis of this paper, namely that the the best possible bandwidth limited linear representation of any object is the convolution of the ground truth object with the ideal OTF, implies that the multiplication with the ideal OTF in eq.~\ref{eq:filtering} is a necessary ingredient of the linear deconvolution.

\begin{figure}[tbhp]
\centering
\includegraphics[width=0.66\linewidth]{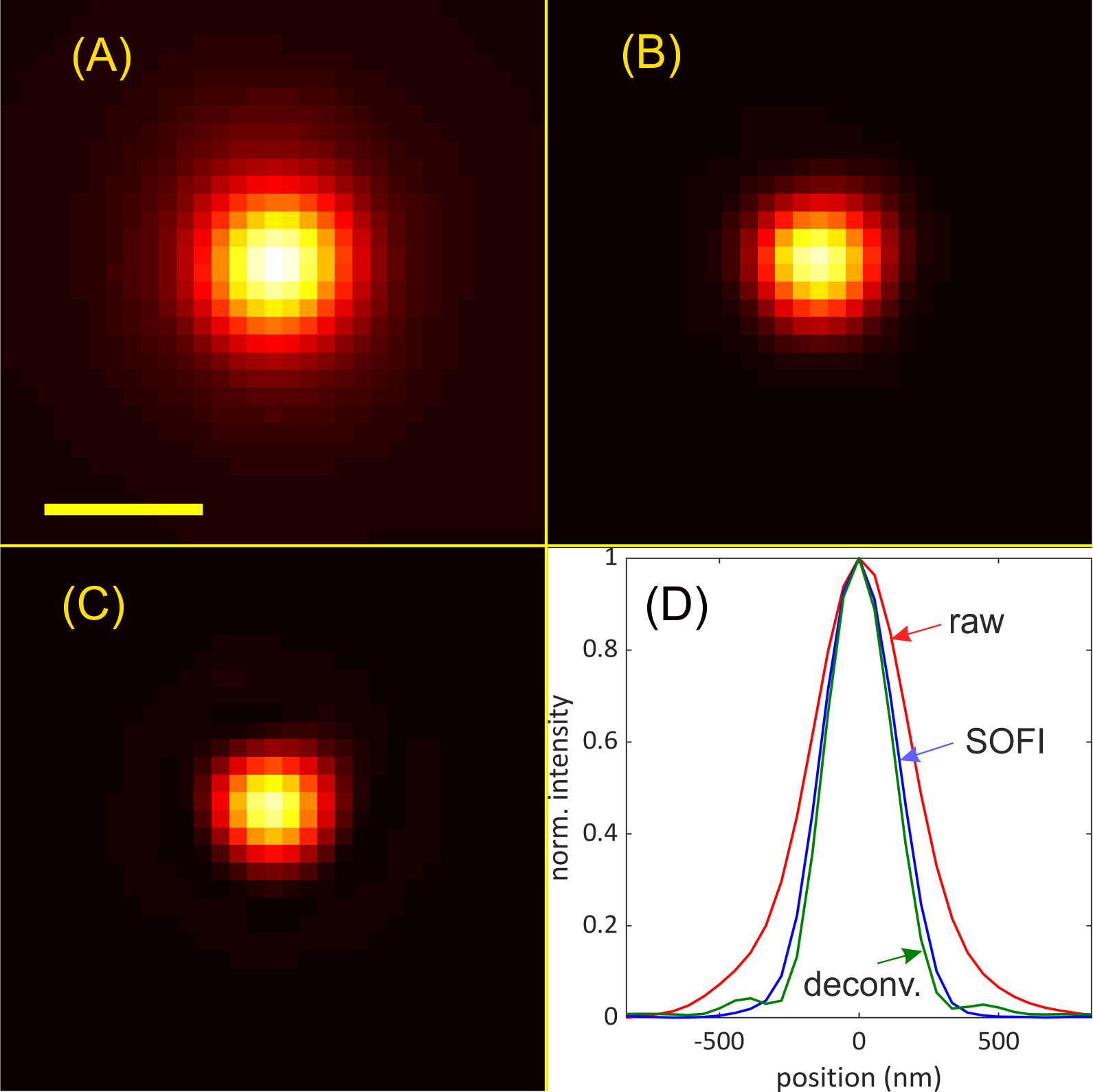}
\caption{SOFI imaging of a single blinking quantum dot on immobilized on a glass cover slide. Panel (A) shows a sum image of a movie of 1000 images, taken with a frame rate 31~Hz on an oil immersion objective with 1.35~NA. Yellow bar is 0.5~µm. Panel (B) shows the second-order SOFI image as calculated following eq.~\ref{eq:SOFI}. Panel (C) shows the deconvolved image using eq.~\ref{eq:filteringeps}, with $\epsilon=0.1 \max\vert\tilde{U}_{exp}(\mathbf{k})\vert$.}
\label{fig:SOFI}
\end{figure}

Of course, a naive scheme such as eq.~\ref{eq:filtering} is numerically dangerous because the denominator $\tilde{U}_{exp}(\mathbf{k})$ tends to zero when the frequency vector $\mathbf{k}$ approaches the boundary of the frequency support $\mathcal{M}$, so one has to apply some regularization to prevent numerical boosting of high-frequency image noise, see e.g. ref.~\cite{muller2010image} for the case of image scanning microscopy (ISM), ref.~\cite{dertinger2010achieving} for the case of super-resolution optical fluctuation imaging (SOFI), or ref.~\cite{smith2021simnoise} for the case of Structured Illumination Microscopy (SIM). In the simplest case, one adds a small number to the denominator to prevent divergence effects, i.e.  
\begin{equation} \begin{split}
\tilde{I}(\mathbf{k}) = \frac{\tilde{U}(\mathbf{k})}{\epsilon + \tilde{U}_{exp}(\mathbf{k})} \tilde{I}_{exp}(\mathbf{k}).
\label{eq:filteringeps}
\end{split} \end{equation}
where $\epsilon$ is typically chosen to be much smaller than the maximum of $\vert\tilde{U}_{exp}(\mathbf{k})\vert$. As an example we have recorded movies of single luminescent quantum dots (QDs) with a wide-field microscope, so that we could apply a Super-resolution Fluorescence Optical Imaging (SOFI) analysis for obtaining a super-resolved second-order SOFI image. In second-order SOFI, one calculates from the movie of recorded images $I(\mathbf{r},t)$ a second-order cumulant image via
\begin{equation} \begin{split}
I_\mathrm{SOFI}(\mathbf{r}) = \sum_{\tau>0} \left< \delta I(\mathbf{r},t) \delta I(\mathbf{r},t+\tau) \right>_t
\label{eq:SOFI}
\end{split} \end{equation}
where $\delta I(\mathbf{r},t) = I(\mathbf{r},t)-\bar{I}(\mathbf{r})$ with $\bar{I}(\mathbf{r})$ being the average (over the whole movie) intensity recorded at position $\mathbf{r}$, the angular brackets denote averaging over time $t$, and cumulants are summed for all possible correlation delay times $\tau$ (for more details, see ref.~\cite{dertinger2010achieving}). The PSF of a second-order SOFI image is equal to the square of the PSF $U_{exp}$ of the original wide-field microscope, or equivalently, its OTF is the auto-convolution of the wide-field OTF. Thus, the ``ideal'' OTF for second order SOFI is the auto-convolution of a uniform amplitude distribution having the same frequency support as that of the wide-field OTF $U_{exp}$. In Fig.~\ref{fig:SOFI} panel (A), the sum image of the movie recorded for a single QD is shown (emission wavelength 670 nm, oil-immersion objective with 1.35 NA, pixel size 55.6~nm), and panel (B) shows the resulting second-order SOFI image. We fitted the wide-field image of the QD (i.e. the wide-field PSF $U_{exp}$) with a scalar approximation of an ideal wide-field 2D-PSF, $U_{exp}(\boldsymbol{\rho}) \propto \left(J_1(\kappa \rho)/\rho\right)^2$, where $J_1$ is the first order Bessel function of the first kind, and $\kappa$ is a fit parameter. This PSF corresponds to an OTF that is the auto-convolution of a uniform amplitude distribution over a disc of spatial frequencies with radius $\kappa$. Thus, this wide-field PSF is ``ideal'' in the sense of our paper, and for the second order SOFI image the ideal PSF is thus given by this fitted PSF scaled down by a factor of two. This corresponds to an up-scaling of the corresponding OTF by the same factor. In panel (C) of Fig.~\ref{fig:SOFI}, we show the resulting image when applying the deconvolution of eq.~\ref{eq:filteringeps} with this up-scaled OTF, an the curves in panel (D) are cross sections for the images shown in panels (A)-(C). When fitting the images with 2D-Gaussian distributions, we indeed find that the three widths of the resulting Gaussians scale ca. as $1:1/\sqrt{2}:1/2$, as expected for second order SOFI and its optimal deconvolution eq.~\ref{eq:filteringeps}.  

Algorithms of more sophistication take into account the frequency-dependence of image noise \cite{stelzer1998contrast,verveer1999comparison,ingerman2019signal}. A recent proposal of one of us for Wiener filtering for SIM \cite{smith2021simnoise} would translate in the current context to finding the Wiener filter $\tilde{W}(\mathbf{k})$ that minimizes \begin{equation} \begin{split}
E = \left< \int d\mathbf{k} \left| \tilde{W}(\mathbf{k})\tilde{I}_{exp}(\mathbf{k}) - \tilde{U}(\mathbf{k})\tilde{S}(\mathbf{k}) \right|^{2} \right>,
\label{eq:wienermetric}
\end{split} \end{equation}
where the angular brackets $\langle\ldots\rangle$ indicate averaging over all noise realizations. That is, we seek a linear deconvolution $\tilde{I}(\mathbf{k})=\tilde{W}(\mathbf{k})\tilde{I}_{exp}(\mathbf{k})$ that best matches the ideal OTF representation of the ground truth object spectrum $\tilde{S}(\mathbf{k})$. Using that
\begin{eqnarray}
\left< \tilde{I}_{exp}(\mathbf{k}) \right> &=& \tilde{U}_{exp}(\mathbf{k})\tilde{S}(\mathbf{k})\\
\left< \left|\tilde{I}_{exp}(\mathbf{k})\right|^{2} \right> &=& 
\left|\tilde{U}_{exp}(\mathbf{k})\right|^{2}\left|\tilde{S}(\mathbf{k})\right|^{2} + \tilde{N}(\mathbf{k}),
\end{eqnarray} 
with $\tilde{N}(\mathbf{k})$ the noise variance in Fourier space, it may be found that the optimum Wiener filtered deconvolution is given by
\begin{equation} \begin{split}
\tilde{I}(\mathbf{k}) = \frac{\tilde{U}(\mathbf{k})}{\tilde{U}_{exp}(\mathbf{k})}
\left[\frac{SSNR(\mathbf{k})}{1 + SSNR(\mathbf{k})}\right]
\tilde{I}_{exp}(\mathbf{k}),
\label{eq:wienerfiltering}
\end{split} \end{equation}
\noindent where $SSNR(\mathbf{k})=\left|\tilde{U}_{exp}(\mathbf{k})\right|^{2}\left|\tilde{S}(\mathbf{k})\right|^{2}/\tilde{N}(\mathbf{k})$ is the Spectral Signal to Noise Ratio. Such an algorithm prevents unwanted boosting of noise, as the $SSNR(\mathbf{k})$ goes to zero when $\mathbf{k}$ approaches the boundary of the frequency support $\mathcal{M}$, just as  the factor $\tilde{U}_{exp}(\mathbf{k})$ in the denominator. 

We have applied this linear deconvolution principle in combination with the proposed ideal OTF concept to a sample of the so-called synaptonemal complex that was imaged both with a Zeiss Airyscan setup, \added{based on the ISM superresolution technique \cite{muller2010image}}, and with a Rescan Confocal Microscopy (RCM, confocal.nl) setup \cite{deluca2013rescan} operating on a Nikon microscope. The underlying image formation of the ISM and RCM techniques are identical, even though the experimental implementations are quite different. In particular, the experimental OTF is found by convolution of the excitation and detection OTF, as explained earlier in this paper. This results in a PSF that effectively is a factor $\sqrt{2}$ more narrow than the PSF of conventional widefield fluorescence microscopy, while the OTF support size is increased by a factor 2. Wiener filtered linear deconvolution as described by eq.~\ref{eq:wienerfiltering} is therefore expected to lead to an additional $\sqrt{2}$ improvement in apparent sharpness.

There were slight differences between the microscopes, in particular the magnification ($M=60$ for Zeiss and $M=150$ for Nikon, using a 1.5$\times$ adapter), Numerical Aperture (NA~=~1.40 for Zeiss and NA~=~1.45 for Nikon) and the back-projected pixel size ($a=42.6$~nm for Zeiss and $a=43.3$~nm for Nikon). The slightly higher NA of the Nikon machine implies that more light is collected for the same exposure time and that the depth of focus is somewhat less. For a sample refractive index $n=1.515$ there is a 13\% smaller depth of focus (this scales proportional to $\lambda/\left(1-\cos\alpha\right)$ with $\sin\alpha=\mathrm{NA}/n$).

The raw data of all 32 detector segments of the Airyscan detector, which are arranged on a hexagonal grid \cite{huff2015airyscan}, were available.
The relative shift of the 32 segments was estimated using the findshift function of DIPlib (www.diplib.org), revealing a spiral arrangement that enables identification of the correct grid position of the segment. 
Then, the images of the 32 segments were added to emulate a confocal image. For the ISM modality we used the found relative shifts to align the 32 images before adding them. The ISM method prescribes an image shift that is 1/2 times the true distance between detector segments. The shift estimation, however, already takes this correctly into account. The estimated distance between detector segments on the hexagonal grid appears to be $2\times(2⁄\sqrt{3})\times a = 98.4$~nm  in sample space, which gives a distance 5.9~µm in the detector plane. The diameter of the entire set of 32 segments is then approximately $6\times 98.4$~nm~=~590~nm, corresponding to 1.59~$\lambda/\mathrm{NA}\approx$~1.30~A.U. which fits reasonably well with practical pinhole diameter settings in confocal microscopy.

\added{The dominant noise source is shot noise governed by Poisson statistics. This implies a noise variance in Fourier space $\tilde{N}(\mathbf{k})$ that is constant and equal to the total (expected) number of collected photons in the image. For the assessment of signal and noise levels it is therefore needed to convert the raw image signals into photon counts.} The raw ISM detector segment images and the RCM images were fed into a routine for estimating the gain and offset \cite{Heintzman2016gain}, giving a gain equal to 4.45 and an offset equal to 0 for the Zeiss system, and a gain of 2.51 and offset of 97.4 for an assumed readout noise of 1.0~e (standard deviation) for the RCM setup, that uses an sCMOS camera (Hamamatsu Orca Flash 4.0). With these parameter estimates the raw data is converted to measured photon counts. It appears that the total number of collected photons is about 4$\times$ higher for the RCM data set compared to the confocal/ISM data set, implying SNR levels that are about 2x higher.

The SSNR is subsequently estimated by averaging over rings in Fourier space \cite{smith2021simnoise}. It appears that this estimate of the SSNR becomes unreliable for the highest spatial frequencies (where essentially the SSNR goes to zero). To fix this, an \emph{ad hoc} normalization:
\begin{equation}
SSNR(\mathbf{k}) \rightarrow \frac{\big|\tilde{U}_{exp}(\mathbf{k})\big|^{2}}{\gamma^{2} + \big|\tilde{U}_{exp}(\mathbf{k})\big|^{2}} SSNR(\mathbf{k})
\label{eq:normalizedssnr}
\end{equation}
for $\gamma=0.05$ is applied, enforcing the physical asymptotic behavior close and beyond the spatial frequency cutoff, without affecting the estimated SSNR for the spatial frequencies within the OTF support.

\begin{figure}[tbhp]
\centering
\includegraphics[width=\linewidth]{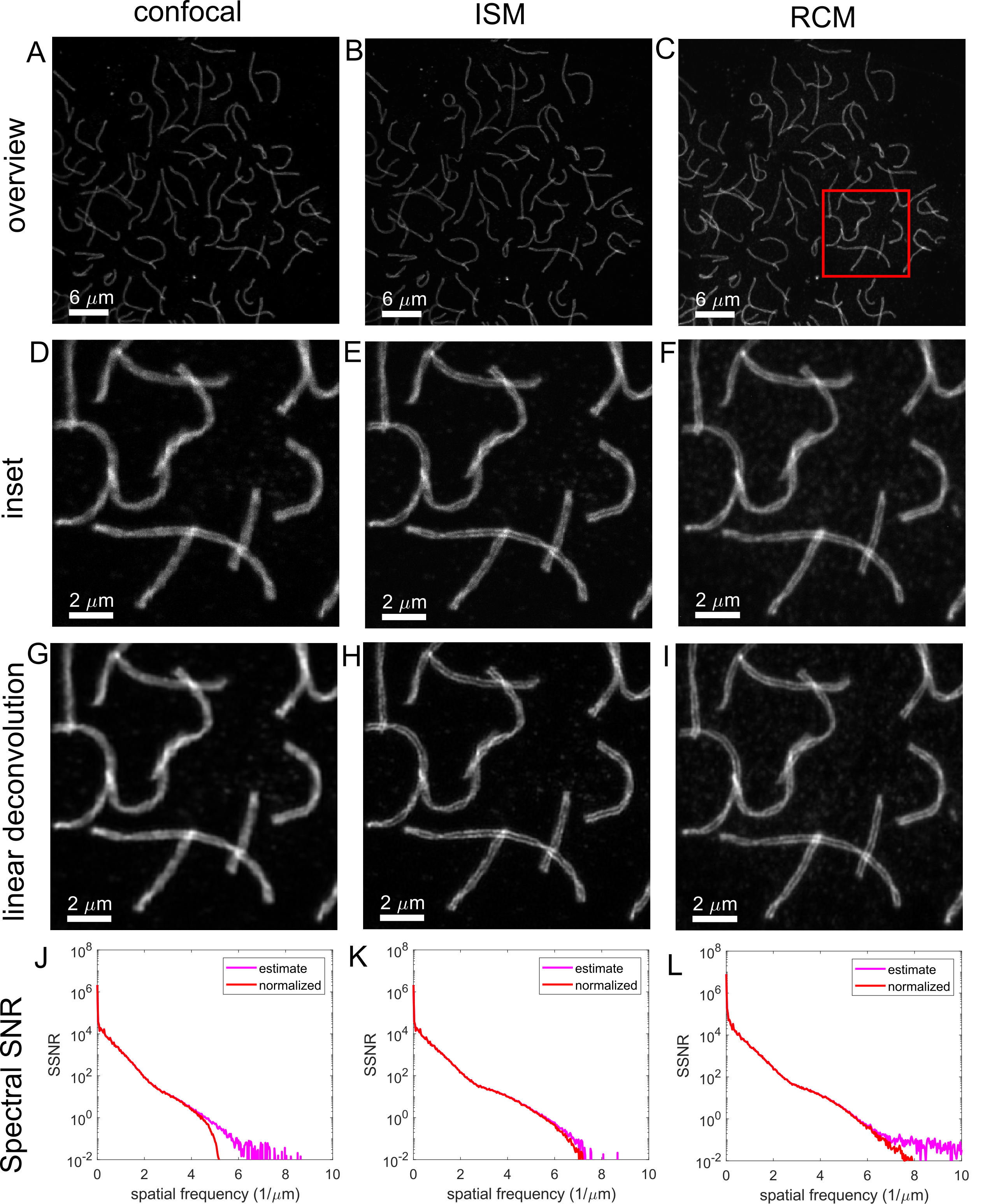}
\caption{Application of the ideal OTF concept in linear deconvolution. Image of a synamptonemal complex sample with confocal (A), ISM (B), and RCM (C) modality. Inset of red box in (C) for the confocal (D), ISM (E), and RCM (F) modality. Linearly deconvolved inset images for the confocal (G), ISM (H), and RCM (I) modality. SSNR estimated by Fourier space ring averaging and normalized SSNR according to Eq.~(\ref{eq:normalizedssnr}) for the confocal (J), ISM (K), and RCM (L) modality.}
\label{fig:RCMvsISM}
\end{figure}

Figure~\ref{fig:RCMvsISM} shows the results of this analysis. The ISM and RCM images before linear deconvolution already show a comparable resolution advantage over the emulated confocal image. Linear deconvolution has a clear impact on all three modalities: noise is suppressed and features appear more clear, in particular the two-line substructure of the filaments can be more clearly recognized in the linearly deconvolved ISM and RCM images. 

\section{Conclusion} 

We have presented a method for obtaining optimum OTFs for a given support in Fourier space and computed the optimum OTF for a range of microscopy modalities. We applied the concept to linear deconvolution of spot scanning microscopy, taking into account the impact of shot noise. There could be alternative or even more general ways to incorporate other noise sources in the treatment, such as  detector thermal noise, read-out noise etc. For example, a possible modification of our algorithm to incorporate noise effects could be to adequately reduce the support of the uniform amplitude distribution which is used for calculating, via auto-convolution, the optimal OTF.

Another possible way to extend the current idea is to explore a numerical, Gerchberg-Saxton-like approach. In each iterative step, the ideal OTF is updated by projecting the Fourier transform of the ideal PSF on the given support $\mathcal{M}$, and the ideal PSF is subsequently updated by taking the absolute value of the Fourier transform of the ideal OTF. Such an approach could be of use if the shape of the OTF support is too complex for an exact direct computation of the ideal OTF, as e.g. for 3D-SIM.

However, the main emphasis of the current paper was to introduce the general idea of an optimal OTF, and to demonstrate it on four idealized examples. Finally, it should be mentioned that our considerations may be valid far beyond the scope of optical microscopy, for example in magnetic resonance imaging, ultrasound imaging, or for deconvolution of telescopic images.

\section{Funding}
JE and AD acknowledge financial support through project ``MICROSCATTAB'' which is jointly financed by a grant of the German Research Foundation (DFG, project number EN297/15-1) and the Agence National de la Recherche (ANR, project number ANR-16-CE92-0003-01). JE thanks also for financial support by the DFG under Germany's Excellence Strategy - EXC 2067/1- 390729940, and by the European Research Council (ERC) via project ``smMIET'' (Grant agreement No. 884488) under the European Union's Horizon 2020 research and innovation programme. \added{NR acknowledges financial by the German Federal Ministry of Education and Research via its funding program ``Photonics Research Germany'' (grant contract number 13N15327).}

\section{Acknowledgments}
We are grateful to Anne Sentenac for proofreading and advice. We thank Jeroen Kole for providing ISM and RCM image data. 

\section{Disclosures}
The authors declare no conflicts of interest.




\end{document}